\documentclass[twocolumn,secnumarabic,amssymb, nobibnotes, aps, pra, superscriptaddress]{revtex4-1}
\usepackage{graphicx}
\usepackage{braket}
\usepackage{array}
\usepackage{amsmath}
\usepackage{here}

\begin{document}

\title{Breaking rotational symmetry in a trapped-ion quantum tunneling rotor}%

\author{Ryutaro Ohira}%
\email{u696585a@ecs.osaka-u.ac.jp}
\affiliation{Graduate School of Engineering Science, Osaka University, 1-3 Machikaneyama, Toyonaka, Osaka, Japan}
\author{Takashi Mukaiyama}%
\affiliation{Graduate School of Engineering Science, Osaka University, 1-3 Machikaneyama, Toyonaka, Osaka, Japan}
\affiliation{Quantum Information and Quantum Biology Division, Institute for Open and Transdisciplinary Research Initiatives, Osaka University, 1-3 Machikaneyama, Toyonaka, Osaka, Japan}
\author{Kenji Toyoda}%
\affiliation{Quantum Information and Quantum Biology Division, Institute for Open and Transdisciplinary Research Initiatives, Osaka University, 1-3 Machikaneyama, Toyonaka, Osaka, Japan}

\date{\today}

\begin{abstract}
A trapped-ion quantum tunneling rotor (QTR) is in a quantum superposition of two different Wigner crystal orientations. In a QTR system, quantum tunneling drives the coherent transition between the two different Wigner crystal orientations. We theoretically study the quantum dynamics of a QTR, particularly when the spin state of one of the ions is flipped. We show that the quantum dynamics of an $\it{N}$-ion QTR can be described by continuous-time cyclic quantum walks. We also investigate the quantum dynamics of the QTR in a magnetic field. Flipping the spin state breaks the rotational symmetry of the QTR, making the quantum-tunneling-induced rotation distinguishable. This symmetry breaking creates coupling between the spin state of the ions and the rotational motion of the QTR, resulting in different quantum tunneling dynamics.
\end{abstract}


\maketitle

\section*{Introduction}

A trapped-ion quantum rigid rotor has recently been realized, and is referred to as a quantum tunneling rotor (QTR) system \cite{1}. Due to the presence of micromotion, it is not straightforward to cool the collective modes of two-dimensional (2D) ion crystals to the motional ground state. However, ground state cooling of the rotational mode, which is a collective mode of a 2D ion crystal, has been demonstrated \cite{1}. The unique property of the QTR is that the rotational motion of the QTR is driven by the quantum tunneling effect. In a previous study, a quantum-tunneling-induced transition between two stable orientations of a Wigner crystal was realized \cite{1}. In addition, quantum interference induced by the Aharonov--Bohm (AB) effect \cite{2} was observed \cite{1}. 

In this paper, we consider the quantum tunneling dynamics of the QTR when the spin state of one of the ions in the QTR is flipped. We first formalize the quantum state of an $\it{N}$-ion QTR including spin degrees of freedom. The quantum dynamics of such a QTR can be considered as a 2$\it{N}$-site cycle graph. Therefore, we describe the quantum dynamics of the QTR as continuous-time cyclic quantum walks \cite{3,4}. Continuous-time cyclic quantum walks require a quantum system where a quantum propagates into neighboring sites via quantum tunneling, and the QTR system is an ideal quantum system for implementing continuous-time cyclic quantum walks.

In addition, we investigate the quantum dynamics of a QTR when a magnetic field passes through the QTR system. Flipping the spin state breaks the rotational symmetry in the QTR system, which then exhibits completely different dynamics. In particular, when the ions in the QTR couple to a vector potential, the difference in the quantum tunneling dynamics becomes marked due to quantum interference induced by the AB effect \cite{1}. We find that this spin-dependent quantum interference induces coupling between the spin state of the ions and the rotational motion of the QTR, which may experimentally realize a quantum spin filter \cite{5} or a cat state of the spin states of the ions and the Wigner crystal structures \cite{6}. In terms of controlling the quantum tunneling dynamics with the spin degrees of freedom of the ions, our work is analogous to studies on the quantum dynamics of a double-well bosonic Josephson junction coupled to a single atomic ion \cite{7,8}.

\begin{figure*}[t] 
\centering
  \includegraphics[width=16.5cm]{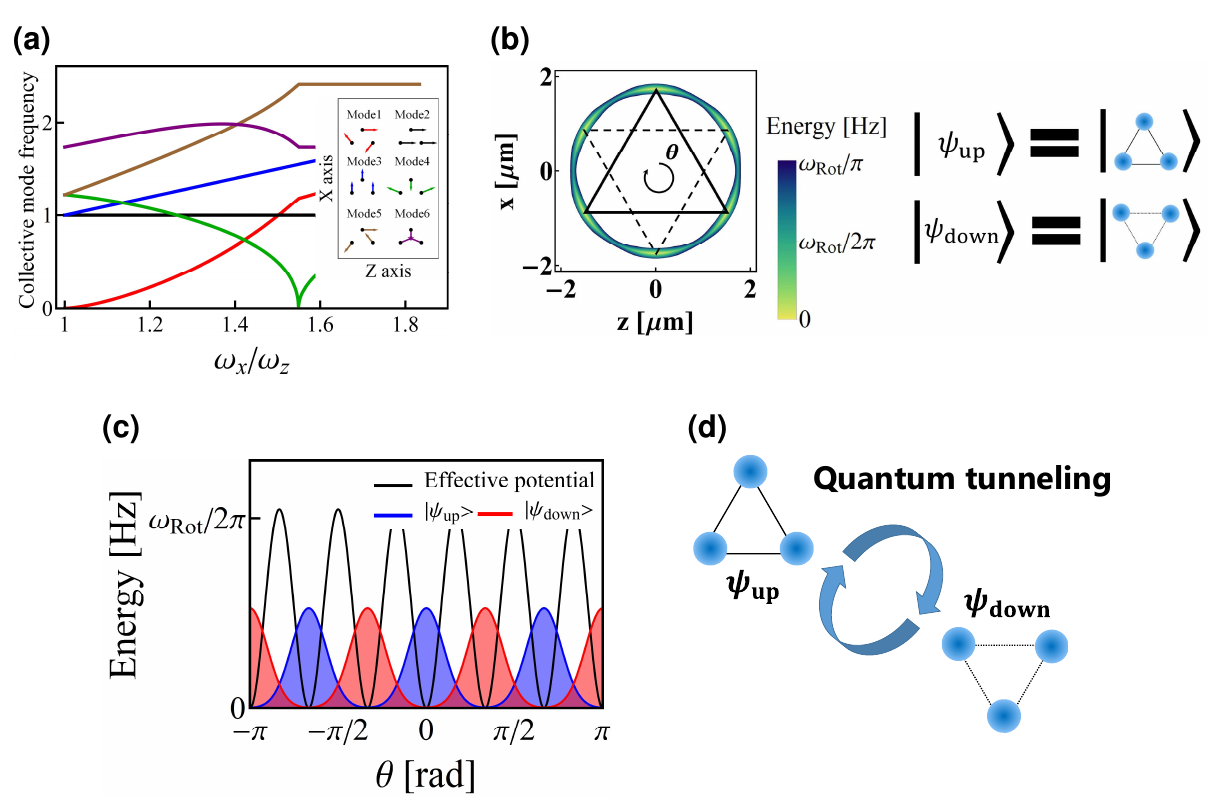}
\caption{\label{fig1}(a) Normalized six lowest collective frequencies of three ions as a function of the ratio of $ \omega_{x} $ to $ \omega_{z} $. The inset shows the eigenvector of each collective mode when $\omega_{x}/\omega_{z}$ = 1.001. (b) Effective potential for $\omega_{x}/\omega_{z}$ = 1.001 ($\omega_{z}$ = 2$\pi\times$1.500 MHz). (c) Effective potential and probability amplitudes of the wavefunctions of $\ket{\psi_{\rm up}}$ and $\ket{\psi_{\rm down}}$ as a function of the angle $\theta$.  (d) Quantum-tunneling-induced transition of the Wigner crystal in the QTR system. }
\end{figure*}

\section*{Trapped-ion QTR}

We first review the dynamics of the QTR. {\it N} ions with mass {\it m} and charge {\it e} are trapped in a harmonic potential. The trap frequencies are  $ \omega_{x} $ and  $ \omega_{z} $ $\ll$ $ \omega_{y} $, so that the motion along the {\it y} direction is considered to be frozen out. The total Hamiltonian of this effective 2D system is ${{\it H}}_{\rm Total} = \sum_{i=1}^{N} {\textit{{p}}}_{i}^2/{2{\it m}}+{\it V}$, where ${\textit{{p}}}_{i}$ represents the momentum of the {\it i}-th ion. The potential {\it V} is derived from the harmonic potential and the Coulomb repulsion; ${\it V} = \sum_{i=1}^{N} {m}(\omega_{x}^2x_{i}^2+\omega_{z}^2z_{i}^2)/2+\sum_{i>j}^{N} {e^2}/{4\pi\varepsilon_{0}\textit{{r}}_{ij}^2}$, where $\textit{{r}}_{ij}$ is the distance between the {\it i}-th and {\it j}-th ions. We assume that the trap frequencies are $ \omega_{z} $$\ll$ $ \omega_{x} $ so that an ion chain along the {\it z} direction is formed. Then, by decreasing the confinement along the {\it x} direction, the ions gradually form a 2D crystal structure. When the trap frequencies $ \omega_{x} $ and $ \omega_{z} $ are comparable, an almost regular polygon crystal is created. 

When the ions have a high mean energy, rotation in the {\it x}--{\it z} plane is observed \cite{1,9}. If the kinetic energy of the ions is lower than the energy barrier for the rotation, the ions are localized at local minima. However, under some conditions, rotation can be driven by the quantum tunneling effect \cite{1}. Specifically, when the motional mean energy is low enough for the ions to be pinned and the rotational barrier is quite low, quantum tunneling occurs. This can be achieved by cooling the rotational mode to the motional ground state and making $ \omega_{x} $ slightly higher than $ \omega_{z} $.

We take a QTR with three trapped ions as an example. We suppose that three $^{171}{\rm Yb}^{+}$ ions are trapped and their internal states are the same. As mentioned above, there are two steps to create a QTR \cite{1}: {\bf1}. Motional ground state cooling of the rotational mode, and {\bf 2}. adiabatically ramping down $ \omega_{x} $ (adiabatic cooling). Figure 1(a) shows a numerical calculation of the frequencies of the six lowest collective modes of the three ions. The red curve (mode1) in Fig.\,1(a) corresponds to the rotational mode. The other collective modes are also shown for reference. In the following discussion, the frequency of the rotational mode is denoted as  $ \omega_{\rm Rot} $, and the trap frequency along the {\it z} direction is fixed to $ \omega_{z}= $ 2$\pi\times$1.500 MHz as a typical experimental parameter. The eigenvectors of each collective modes for $\omega_{x}/\omega_{z}$ = 1.001 are also shown in Fig.\,1(a). 

To create the QTR, the rotational mode needs to be cooled to the motional ground state; otherwise the ions start rotating due to the high kinetic energy. Therefore, we first set $ \omega_{x} $ to the point where $ \omega_{\rm Rot} $ is high enough to cool the rotational mode to the ground state. After motional ground state cooling, adiabatic cooling \cite{1,10,11,12} is employed to increase the population of the ground state of the rotational mode. As indicated in Fig.\,1(a), $ \omega_{\rm Rot} $ gets smaller as $ \omega_{x} $ is ramped down. By reducing the confinement along the {\it x} direction under the condition $\frac{d\omega_{\rm Rot}}{dt}$/$ \omega_{\rm Rot} $ $\ll$ 1, adiabatic cooling is realized. $ \omega_{x} $ is adiabatically ramped down to the point where the trap frequencies along the {\it x} and {\it z} directions are comparable. In this article, we discuss the dynamics of a QTR assuming perfect ground state cooling of the rotational mode, so that the classical rotation caused by the other quantum motional states, which have the higher kinetic energy than the effective rotational barrier, can be safely ignored. Figure 1(b) shows the effective potential created by the harmonic potential and the Coulomb interaction between ions for $\omega_{x}/\omega_{z}$ = 1.001. It is clear from Fig.\,1(b) that there are two stable ion crystal orientations. We define the wavefunctions of those stable ion structures as $\ket{\psi_{\rm up}}$ and $\ket{\psi_{\rm down}}$.

Figure 1(c) shows the effective potential and the probability amplitudes for both wavefunctions, $\ket{\psi_{\rm up}}$ and $\ket{\psi_{\rm down}}$, for $\omega_{x}/\omega_{z}$ = 1.001. As can be seen from Fig.\,1(c), the wavefunctions $\ket{\psi_{\rm up}}$ and $\ket{\psi_{\rm down}}$ overlap. Since $\ket{\psi_{\rm up}}$ and $\ket{\psi_{\rm down}}$ are not orthogonal, quantum tunneling can occur.

The QTR changes its crystal orientation due to the quantum tunneling effect, as shown in Fig.\,1(d). The Coulomb interaction prevents the exchange of ions when quantum tunneling occurs. Therefore, the QTR can be considered to be a quantum rigid rotor system. It should be noted that this protocol to create the QTR is not realistic if the number of ions is even because the barrier for rotation is significantly high.

\section*{Cyclic quantum walks}

\subsection*{Formation of QTR, including the spin state}

An ion has an internal state, which is considered an effective spin state. For $^{171}{\rm Yb}^{+}$ ions, the hyperfine states $\ket{\downarrow}$ $\equiv$ $\ket{{\it F}=0, {\it m}_{F}=0}$ and $\ket{\uparrow}$ $\equiv$ $\ket{{\it F}=1, {\it m}_{F}=0}$ are usually used as the effective spin system. If the internal states of the ions are the same, $\ket{\downarrow}$, the internal states are simply ignored and the quantum state of the {\it N}-ion QTR, $\ket{\psi}$, can be expressed as the superposition of $\ket{\psi_{\rm up}}$ and $\ket{\psi_{\rm down}}$: 
\begin{equation}
  \ket{\psi} = \alpha\ket{\psi_{\rm up}}+\beta\ket{\psi_{\rm down}}, 
\end{equation}
where $\alpha$ and $\beta$ are complex coefficients satisfying $|\alpha|^2$+$|\beta|^2$ = 1. Since the QTR is the superposition of two different polygon Wigner crystals, the {\it N}-ion QTR is viewed as a 2{\it N}-site cycle graph, as shown in Fig.\,2. We define the wavefunction of the {\it k}-th ion at the {\it l}-th site as $\ket{\psi_{k,l}}$. As is evident from quantum interference in the QTR system induced by the AB effect \cite{1}, the particles constituting the QTR are considered to be identical. Since $^{171}{\rm Yb}^{+}$ is bosonic, the quantum states of $\ket{\psi_{\rm up}}$ and $\ket{\psi_{\rm down}}$ are given as follows:
\allowdisplaybreaks
\begin{gather}
\ket{\psi_{\rm up}} = \frac{1}{\sqrt{N!}} \nonumber \\
{\rm perm}\left(
\begin{array}{ccccc}
  \ket{\psi_{1,1}} & \cdots & \ket{\psi_{1,2n-1}} & \cdots & \ket{\psi_{1,2N-1}} \\
  \vdots & \ddots & \vdots & \ddots & \vdots \\
  \ket{\psi_{k,1}} & \cdots & \ket{\psi_{k,2n-1}} & \cdots & \ket{\psi_{k,2N-1}} \\
  \vdots & \ddots & \vdots & \ddots & \vdots \\
  \ket{\psi_{N,1}} & \cdots & \ket{\psi_{N,2n-1}} & \cdots & \ket{\psi_{N,2N-1}}
\end{array}
\right),
\end{gather}
\allowdisplaybreaks
\begin{gather}
\ket{\psi_{\rm down}} = \frac{1}{\sqrt{N!}} \nonumber \\
{\rm perm}\left(
\begin{array}{ccccc}
  \ket{\psi_{1,2}} & \cdots & \ket{\psi_{1,2n}} & \cdots & \ket{\psi_{1,2N}} \\
  \vdots & \ddots & \vdots & \ddots & \vdots \\
  \ket{\psi_{k,2}} & \cdots & \ket{\psi_{k,2n}} & \cdots & \ket{\psi_{k,2N}} \\
  \vdots & \ddots & \vdots & \ddots & \vdots \\
  \ket{\psi_{N,2}} & \cdots & \ket{\psi_{N,2n}} & \cdots & \ket{\psi_{N,2N}}
\end{array}
\right).
\end{gather}
When the ions are fermionic, $\ket{\psi_{\rm up}}$ and $\ket{\psi_{\rm down}}$ can be expressed using Slater determinants. There are two directions of the rotations induced by quantum tunneling in the QTR system. Therefore, the Hamiltonian of the QTR is given as follows:
\begin{figure}[t]
\centering
  \includegraphics[width=8cm]{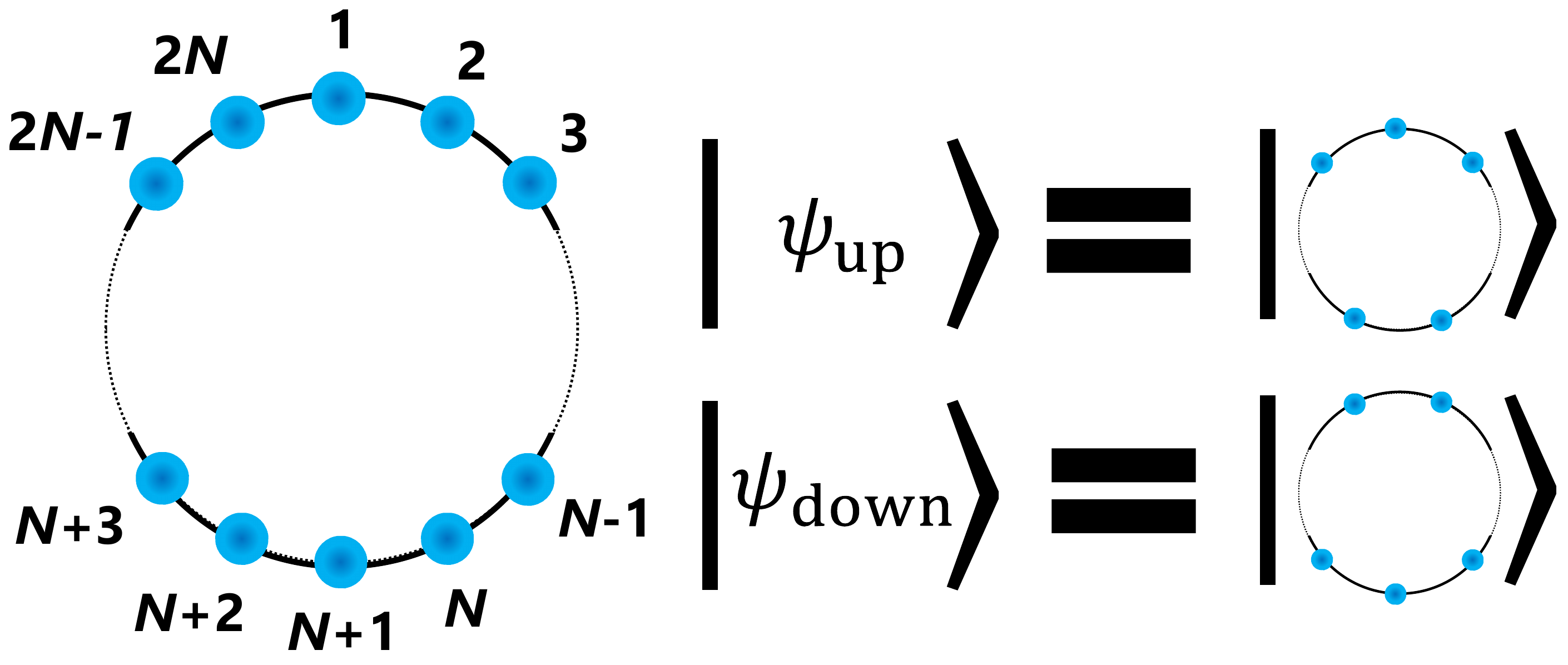}
\caption{\label{fig1}Schematic representation of the {\it N}-ion QTR. Since the QTR is the superposition of two different polygon Wigner crystal orientations $\ket{\psi_{\rm up}}$ and $\ket{\psi_{\rm down}}$, the QTR is considered to be a 2{\it N}-site cycle graph.}
\end{figure}
\begin{equation}
  {\it H} = \hbar{\it j}({\hat{J}}_{\rm CL}+{\hat{J}}_{\rm CCL}),
\end{equation}
where $\hbar$ and ${\it j}$ are the Planck constant and the quantum tunneling rate. ${\hat{J}}_{\rm CL}$ and ${\hat{J}}_{\rm CCL}$ are defined as 
\begin{equation}
  {\hat{J}}_{\rm CL}\equiv\bigotimes_{k=1}^{N}\sum_{l=1}^{2N}\ket{\psi_{k,l+1}}\bra{\psi_{k,l}}, 
\end{equation}
\begin{equation}
{\hat{J}}_{\rm CCL}\equiv\bigotimes_{k=1}^{N}\sum_{l=1}^{2N}\ket{\psi_{k,l}}\bra{\psi_{k,l+1}},
\end{equation}
with $\ket{\psi_{k,2N+1}}=\ket{\psi_{k,1}}$. Thus, ${\hat{J}}_{\rm CL}$ and ${\hat{J}}_{\rm CCL}$ satisfy ${\hat{J}}_{\rm CL}\ket{\psi_{\rm up(down)}}=\ket{\psi_{\rm down(up)}}$ and ${\hat{J}}_{\rm CCL}\ket{\psi_{\rm down(up)}}=\ket{\psi_{\rm up(down)}}$, representing clockwise and counter-clockwise rotation, respectively. It is possible to experimentally distinguish these two states using a projective measurement \cite{1}, i.e., there are two distinguishable states. 

Next, we flip the internal state of the ion at the {\it l}-th site ({\it l} = 2{\it n} or 2{\it n}-1). By introducing the wavefunction
\allowdisplaybreaks
\begin{gather}
\ket{\psi_{\rm up,2n-1}} = \frac{1}{\sqrt{N!}} \nonumber \\
{\rm perm}\left(
\begin{array}{ccccc}
  \ket{\downarrow}\ket{\psi_{1,1}} & \cdots & \ket{\uparrow}\ket{\psi_{1,2n-1}} & \cdots & \ket{\downarrow}\ket{\psi_{1,2N-1}} \\
  \vdots & \ddots & \vdots & \ddots & \vdots \\
  \ket{\downarrow}\ket{\psi_{k,1}} & \cdots & \ket{\uparrow}\ket{\psi_{k,2n-1}} & \cdots & \ket{\downarrow}\ket{\psi_{k,2N-1}} \\
  \vdots & \ddots & \vdots & \ddots & \vdots \\
  \ket{\downarrow}\ket{\psi_{N,1}} & \cdots & \ket{\uparrow}\ket{\psi_{N,2n-1}} & \cdots & \ket{\downarrow}\ket{\psi_{N,2N-1}}
\end{array}
\right)
\end{gather}
and
\allowdisplaybreaks
\begin{gather}
\ket{\psi_{\rm down,2n}} = \frac{1}{\sqrt{N!}} \nonumber \\
{\rm perm}\left(
\begin{array}{ccccc}
  \ket{\downarrow}\ket{\psi_{1,2}} & \cdots & \ket{\uparrow}\ket{\psi_{1,2n}} & \cdots & \ket{\downarrow}\ket{\psi_{1,2N}} \\
  \vdots & \ddots & \vdots & \ddots & \vdots \\
  \ket{\downarrow}\ket{\psi_{k,2}} & \cdots & \ket{\uparrow}\ket{\psi_{k,2n}} & \cdots & \ket{\downarrow}\ket{\psi_{k,2N}} \\
  \vdots & \ddots & \vdots & \ddots & \vdots \\
  \ket{\downarrow}\ket{\psi_{N,2}} & \cdots & \ket{\uparrow}\ket{\psi_{N,2n}} & \cdots & \ket{\downarrow}\ket{\psi_{N,2N}}
\end{array}
\right),
\end{gather}
including the spin degrees of freedom, the quantum state of the QTR can be given as follows:

\begin{figure}[t] 
\centering
  \includegraphics[width=8.5cm]{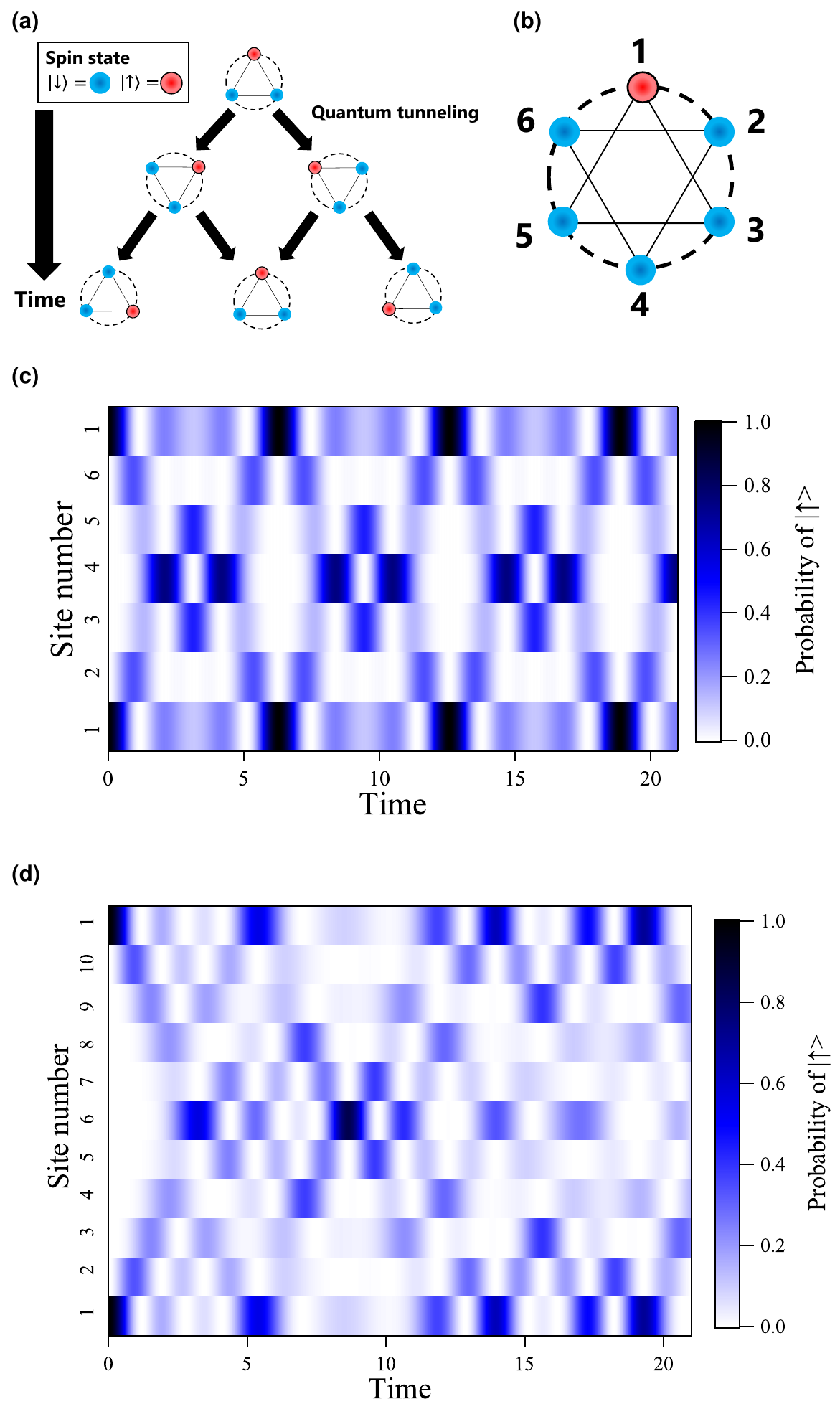}
\caption{\label{fig1} (a) Time evolution of the 3-ion QTR. The spin states $\ket{\uparrow}$ and $\ket{\downarrow}$ are represented as red and blue circles, respectively. (b) Time evolution of the QTR shown in Fig.\,3(a), considered as a 6-site cyclic graph. The ion in $\ket{\uparrow}$ propagates into the neighboring sites via quantum tunneling. (c,d) Continuous-time cyclic quantum walks. The time evolution of the probability distribution of the ion in $\ket{\uparrow}$ is calculated for (c) 3-ion and (d) 5-ion QTR.}
\end{figure}

\begin{equation}
\ket{\psi} = \alpha'\sum_{n=1}^{N} {c_{2n-1}\ket{\psi_{\rm up,2n-1}}}+\beta'\sum_{n=1}^{N} {c_{2n}\ket{\psi_{\rm down,2n}}},
\end{equation}
where $\alpha'$, $\beta'$, $c_{2n-1}$ and $c_{2n}$ are the complex coefficients satisfying $|\alpha'|^2$+$|\beta'|^2$ = 1, $\sum_{n=1}^{N} {|c_{2n-1}|^2}=1$ and $\sum_{n=1}^{N} {|c_{2n}|^2}=1$. This equation clearly shows that there are distinguishable {\it N} states for each crystal orientation. Therefore, a QTR with an ion in the internal state $\ket{\uparrow}$ at the {\it l}-th site is considered to be a 2{\it N}-site cyclic graph. Figure\,3(a) shows the time evolution of a 3-ion QTR with one ion in $\ket{\uparrow}$. As is clear from the above discussion, as shown in Fig.\,3(b), this model is considered to be equivalent to the 2{\it N}-site cycle graph where the ion with internal state $\ket{\uparrow}$ at the {\it n}-th site evolves into the neighboring sites via quantum tunneling. By introducing the notation $\ket{2n-1}=\ket{\psi_{\rm up,2n-1}}$ and $\ket{2n}=\ket{\psi_{\rm down,2n}}$, the quantum state of the QTR can be rewritten as 
\begin{equation}
\ket{\psi} = \sum_{n=1}^{2N} {\gamma_{n}\ket{n}},
\end{equation}
where $\gamma_{n}$ is the complex coefficient satisfying $\sum_{n=1}^{2N} {|\gamma_{n}|^2}=1$. The Hamiltonian of this system can also be described as
\begin{equation}
  {\it H} = \hbar{\it j}\sum_{n=1}^{2N}(\hat{a}_{n}\hat{a}_{n+1}^{\dagger}+\hat{a}_{n}^{\dagger}\hat{a}_{n+1}),
\end{equation}
where $\hat{a}_{1}=\hat{a}_{2N+1}$. $\hat{a}_{n}^{\dagger}$ and $\hat{a}_{n}$ are creation and annihilation operators of an ion in $\ket{\uparrow}$ at the {\it n}-th site.

We show the quantum dynamics of the 3-ion QTR in Fig.\,3(c). In the simulation, we prepared the initial state in $\ket{1}$, as shown in Fig.\,3(b). Note that the evolution time is normalized to the unit of the quantum tunneling rate. Here, by considering the ion in $\ket{\uparrow}$ to be a quantum walker, the time evolution of the probability distribution of the ion in $\ket{\uparrow}$ can be described by continuous-time quantum walks. The quantum dynamics of the cyclic quantum walks using a 5-ion QTR is also shown in Fig.\,3(d) (see appendix for more details).

\section*{Spin-dependent\\quantum interference}

We now consider a magnetic flux $\Phi=\it{S}\it{B}$ threading through a QTR consisting of identical ions, as shown in Fig.\,4(a). Here, $\it S$ and $\it B$ are a closed area of the QTR and a magnetic field passing through the closed area, respectively. The ions in the QTR couple to the vector potential. Therefore, the AB effect introduces a relative phase difference $2\theta_{\rm AB}$ between wavefunctions rotating clockwise and counterclockwise, as shown in Fig.\,4(b) \cite{1}. The AB effect induced phase shift is $\theta_{\rm AB}=\pi\frac{\Phi}{\phi_{\rm 0}}$, where ${\phi_{\rm 0}}=\frac{\hbar}{e}$. Here, the Hamiltonian of the QTR is expressed as 
\begin{equation}
  {\it H} = \hbar{\it j}({\hat{J}}_{\rm CL}e^{i\theta_{\rm AB}}+{\hat{J}}_{\rm CCL}e^{-i\theta_{\rm AB}}).
\end{equation}

Fig.\,4(c) shows the numerically calculated time evolution of the probability of finding $\ket{\psi_{\rm up}}$ as a function of time. In the numerical simulation, we prepared an initial state of the 3-ion QTR in $\ket{\psi_{\rm up}}$. Then, we calculated the probability of finding $\ket{\psi_{\rm up}}$ based on Eq.\,(12). Since the spin states of the ions in the QTR are the same, the rotational direction of the transition of the QTR is indistinguishable. Therefore, the quantum interference is induced by the AB effect \cite{1}. When the AB effect phase shift is $\pi/2$, the clockwise and counter-clockwise rotating wavefunctions counteract each other, resulting in suppression of quantum tunneling. These results clearly show that it is possible to control the quantum tunneling probability by changing the amount of magnetic flux passing through the QTR. By combining the quantum interference effect and the time-dependent magnetic field, it is possible to realize arbitrary superposition of two different Wigner crystal orientations, such as $\ket{\psi}=\frac{1}{\sqrt{2}}(\ket{\psi_{\rm up}}+e^{i\theta_{\rm 0}}\ket{\psi_{\rm down}})$, where $\theta_{\rm 0}$ is the relative phase shift.

We next consider the quantum dynamics of a QTR for which one of the spin states of the ions is flipped. According to the above discussion and Eq.\,(11), the Hamiltonian of this QTR system is described as follows:
\begin{equation}
  {\it H} = \hbar{\it j}\sum_{n=1}^{2N}(\hat{a}_{n}\hat{a}_{n+1}^{\dagger}e^{i\theta_{\rm AB}}+\hat{a}_{n}^{\dagger}\hat{a}_{n+1}e^{-i\theta_{\rm AB}}),
\end{equation}

\begin{figure}[t]
\centering
  \includegraphics[width=8.5cm]{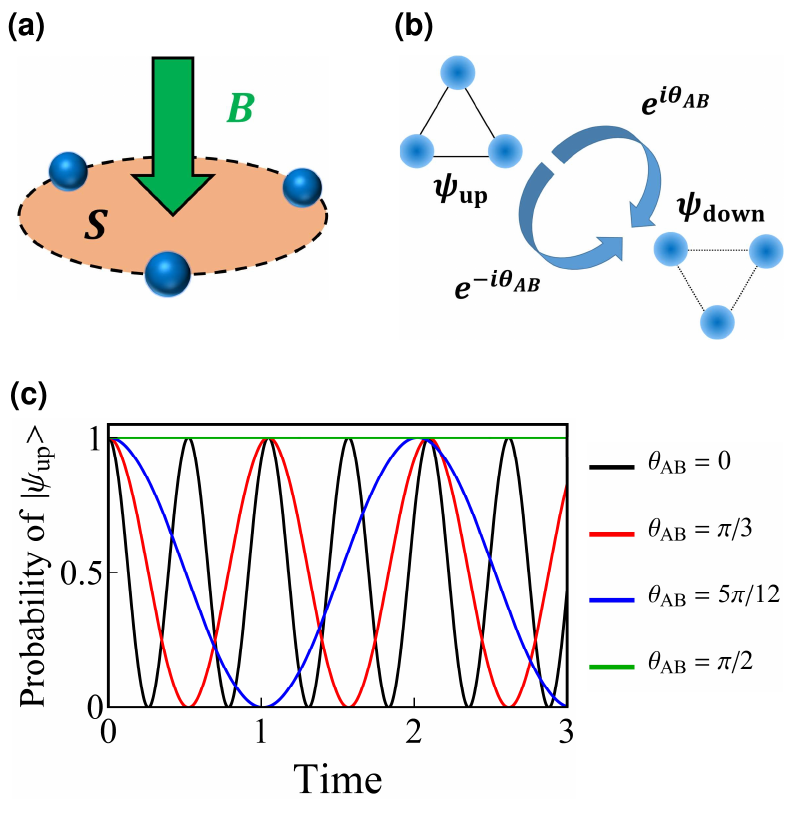}
\caption{\label{fig4} (a) Magnetic field $\it{B}$ passing through the 3-ion QTR. (b) Ions in the QTR system coupled to the vector potential, introducing a phase difference between wavefunctions rotating in two opposite directions. (c) Probability of finding $\ket{\psi_{\rm up}}$ for each AB phase shift as a function of time. The evolution time is normalized to the unit of the quantum tunneling rate.}
\end{figure}

\begin{figure}[t]
\centering
  \includegraphics[width=7.9cm]{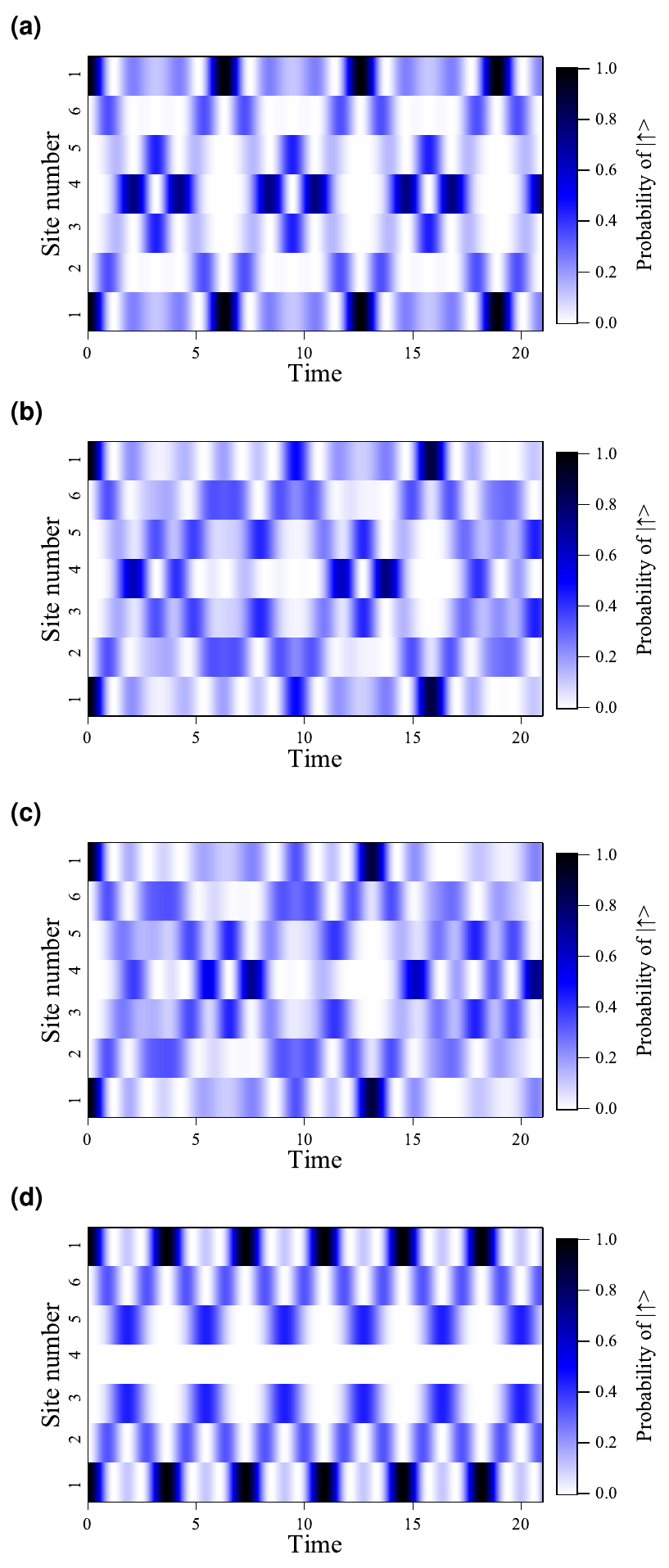}
\caption{\label{fig5} (a--d) Time evolution of the probability distribution of an ion in $\ket{\uparrow}$. The initial state is prepared in $\ket{1}$. The evolution time is normalized to the unit of the quantum tunneling rate. (a) $\theta_{\rm AB}=0$. (b) $\theta_{\rm AB}=\frac{\pi}{24}$. (c) $\theta_{\rm AB}=\frac{\pi}{12}$. (d) $\theta_{\rm AB}=\frac{\pi}{6}$.}
\end{figure}

We show a simulation of the 3-ion QTR dynamics when the AB phase shifts are $\theta_{\rm AB}=0, \frac{\pi}{24}, \frac{\pi}{12}$, and $\frac{\pi}{6}$ in Fig.\,5(a--d). Note that the time is normalized to the unit of the quantum tunneling rate and the initial state is prepared in $\ket{1}$. The dynamics are slightly modulated by the corresponding $\theta_{\rm AB}$. In particular, when $\theta_{\rm AB}=\frac{\pi}{6}$, destructive interference always occurs at the opposite site, i.e., at 4-th site, resulting in the probability of finding $\ket{4}$ being always zero, as shown in Fig.\,5(d).

We find that the spin degrees of the trapped ions in a QTR can change the quantum tunneling dynamics of the QTR. This property can be used to realize spin-dependent motion of the QTR. For example, we consider the dynamics of a 3-ion QTR having identical ions. By applying the magnetic field which induces $\theta_{\rm AB}=\frac{\pi}{2}$, the QTR does not rotate due to the destructive interference, as shown in Fig.\,4(c). However, if we flip the spin state of one of the ions in the QTR, the QTR then starts to rotate, as shown in Fig.\,5(d). The dynamics of the QTR with an ion in $\ket{\uparrow}$ when $\theta_{\rm AB}=\frac{\pi}{2}$ is exactly the same as Fig.\,5(d). This is analogous to spin-motion coupling, such as the Stern--Gerlach experiment \cite{13} or the quantum spin filter \cite{5}. In addition, if one of the ions is prepared in $\frac{1}{\sqrt{2}}(\ket{\uparrow}+\ket{\downarrow})$, the superposition of two different QTRs is realized, for which one QTR rotates and the another does not. Such a quantum system may realize the entangled state of the spin degrees of freedom and the Wigner crystal structures \cite{6}.

\section*{Discussion}

To observe quantum tunneling, the quantum coherence of the QTR is important. One of the biggest sources of decoherence is heating of the rotational mode. The quantum tunneling rate of the QTR is relatively slow. For example, the numerically calculated quantum tunneling rate using the wavefunctions and effective potential shown in Fig.\,1(c) is 4.95 Hz. Therefore, the heating rate of the rotational mode needs to be sufficiently suppressed. As previously discussed \cite{1}, the lack of adiabaticity of the trap potential control and the fluctuation of the RF voltage may be related to the heating of the rotational mode. These factors can be avoided by carefully choosing the experimental parameters and using the RF voltage stabilization method \cite{14}. Other collective modes may heat the rotational mode. To prevent this effect, further implementation of the laser cooling technique to cool all the collective modes simultaneously \cite{15} may be helpful. Heating of the rotational mode decreases the population in the motional ground state of the rotational mode, reducing the quantum tunneling quality. However, as long as the population of the motional ground state of the rotational mode is not zero, the quantum tunneling dynamics can be deduced from the measurement results. In addition, the coherence time of the quantum tunneling needs to be long. One reason for the degradation of the quantum tunneling coherence is the fluctuation of the RF potential. This can be mitigated by implementing RF stabilization \cite{14}.

\section*{Conclusions}
We have studied the quantum dynamics of a QTR when a spin state of one of the ions is flipped. We found that symmetry breaking by flipping the spin state of the ion of the QTR results in different quantum dynamics. Additionally, we have exploited the quantum dynamics of the QTR when a magnetic field is present. Our work may be useful for realizing applications using spin-motion coupling \cite{5,6,7,8}. While quantum tunneling is a fundamental phenomenon, it is difficult to investigate the quantum tunneling dynamics at a single quantum level. Advantages such as individual addressability and manipulation of quantum states of the ions provide a well-controllable quantum system. Therefore, a trapped-ion QTR system can provide an ideal platform for performing fundamental quantum physics experiments.

\section*{Acknowledgments}
R.O. would like to thank Go Tomimasu and Atsushi Noguchi for fruitful discussion. This work was supported by MEXT Quantum Leap Flagship Program (MEXT Q-LEAP) Grant Number JPMXS0118067477.

\begin{figure*}[t]
\centering
  \includegraphics[width=16cm]{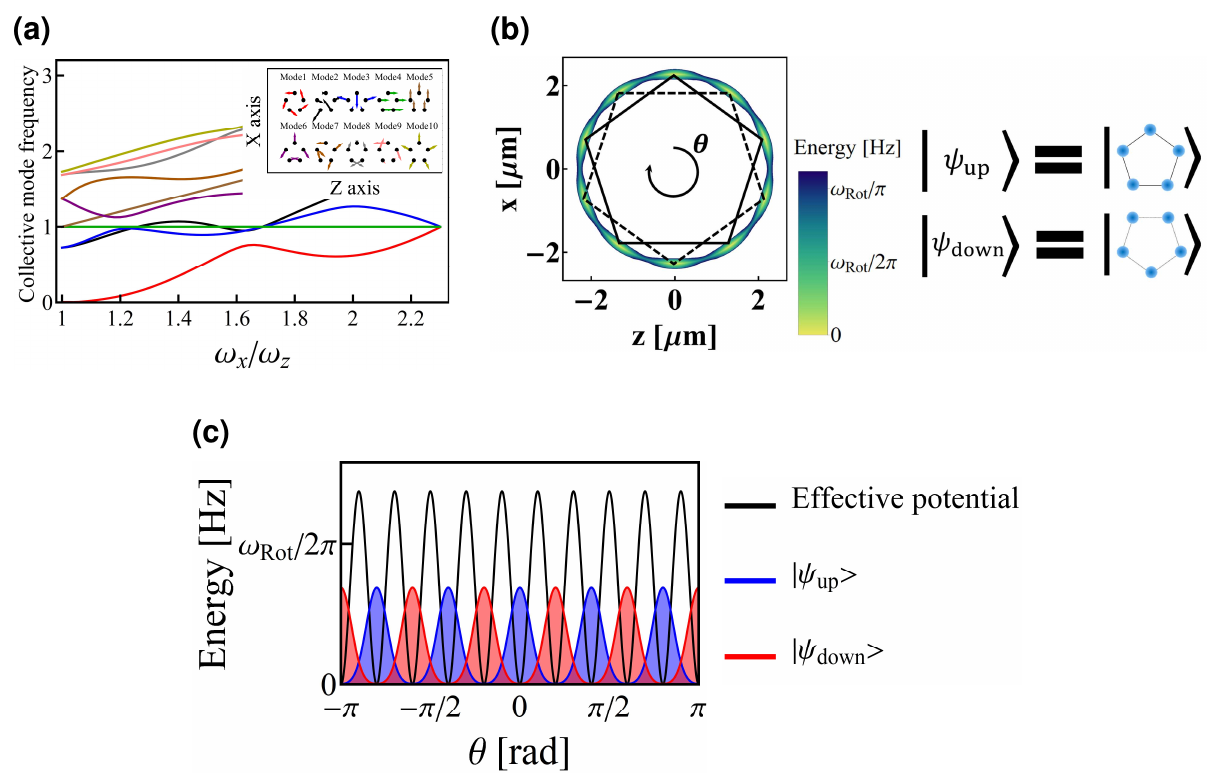}
\caption{\label{fig6}(a) Normalized ten lowest collective frequencies of the five trapped ions as a function of the ratio of $ \omega_{x} $ to $ \omega_{z} $. The inset shows the eigenvector of each collective mode for  $\omega_{x}/\omega_{z}$ = 1.010. (b) Effective potential for $\omega_{x}/\omega_{z}$ = 1.010 ($\omega_{z}$ = 2$\pi\times$1.500 MHz). The almost isotropic trap potential and Coulomb interaction create two stable ion crystal orientations $\ket{\psi_{\rm up}}$ and $\ket{\psi_{\rm down}}$. (c) Effective potential and amplitudes of the wavefunctions $\ket{\psi_{\rm up}}$ and $\ket{\psi_{\rm down}}$ as a function of the angle $\theta$.}
\end{figure*}

\appendix* \section{5-ion QTR}
We show the results of the analysis of a QTR using five $^{171}{\rm Yb}^{+}$ ions. We assume that the confinement along the {\it z} direction ($ \omega_{z} $) is 2$\pi\times$1.500 MHz. The numerically calculated frequencies of the ten lowest collective modes are shown in Fig.\,6(a). The eigenvectors of each collective mode for $\omega_{x}/\omega_{z}=1.010$ are also shown in Fig.\,6(a). The red curve (mode1) is the rotational mode. Fig.\,6(b) shows the effective potential of the 5-ion QTR system for $\omega_{x}/\omega_{z}$ = 1.010. There are two stable orientations of the ion crystal, $\ket{\psi_{\rm up}}$ and $\ket{\psi_{\rm down}}$. When $\omega_{x}/\omega_{z}$ = 1.010, $\ket{\psi_{\rm up}}$ and $\ket{\psi_{\rm down}}$ overlap [Fig.\,6(c)].

\end{document}